\documentclass[aps,prd,twocolumn,nofootinbib,superscriptaddress]{revtex4}
\usepackage{latexsym}
\usepackage{hyperref}
\usepackage{amssymb}
\usepackage{epsfig,amsmath,graphics}
\usepackage{pstricks}
\usepackage{color}
\usepackage{dcolumn}
\usepackage{ulem}
\usepackage{xcolor}
\usepackage{placeins}
\usepackage{siunitx}
\usepackage{float}

\definecolor{linkcolor}{rgb}{0.0, 0.28, 0.67}
\hypersetup{colorlinks=true,
linkcolor=black,
citecolor=linkcolor,
urlcolor=linkcolor,
linktocpage=true
}

\newcommand{\Neff}{N_{\rm eff}}

\newcommand{\NIR}{N_{\rm int}^{\rm IR}}

\newcommand{\LCDM}{\Lambda{\rm CDM}}

\newcommand{\TDS}{T_{\rm DS}}

\newcommand{\LL}{\mathcal{L}}
\newcommand{\HH}{\mathcal{H}}

\newcommand{\DD}{\mathcal{D}}

\newcommand{\nir}{N_\text{IR}}

\usepackage{comment}
\newcommand{\be}{\begin{equation}}
\newcommand{\ee}{\end{equation}}

\def\bea{\begin{eqnarray}}
\def\eea{\end{eqnarray}}
\def\ltap{\ \raise.3ex\hbox{$<$\kern-.75em\lower1ex\hbox{$\sim$}}\ }
\def\gtap{\ \raise.3ex\hbox{$>$\kern-.75em\lower1ex\hbox{$\sim$}}\ }
\def\lsim{\ \raise.3ex\hbox{$<$\kern-.75em\lower1ex\hbox{$\sim$}}\ }
\def\gsim{\ \raise.3ex\hbox{$>$\kern-.75em\lower1ex\hbox{$\sim$}}\ }

\usepackage{slashed}

\usepackage{color}

\newcommand{\ignore}[1]{}
\newcommand{\beq}{\begin{equation}}
\newcommand{\eeq}{\end{equation}}
\newcommand{\bear}{\begin{eqnarray}}
\newcommand{\eear}{\end{eqnarray}}

\def\MeV{\,{\rm MeV}}

\def\lcdm{$\Lambda{\rm CDM}$}

\begin{document}

\title{Searching for Dark Matter Interactions with ACT, SPT and DES}

\author{Zilu Zhou}
\affiliation{Center for Cosmology and Particle Physics, Department of Physics, New York University, New York, NY 10003, USA}
\author{Neal Weiner}
\affiliation{Center for Cosmology and Particle Physics, Department of Physics, New York University, New York, NY 10003, USA}
\begin{abstract}
Models of a dark radiation sector with a mass threshold (WZDR+) have proved to be an appealing alternative to $\LCDM$. These models provide simple comparison models, grounded in well-understood particle physics and with limited additional parameters. In addition, they have shown relevance in easing existing cosmological tensions, specifically the $H_0$ tension and the $S_8$ tension. Recently, measurements of CMB lensing by the ACT collaboration have provided strong additional information on clustering at late times. Within $\LCDM$, these results yield a high value of $S_8$ at odds with weak-lensing measurements. In this work, we study this in the context of WZDR+, and find a much wider range of allowed values of $S_8$, and in particular much better agreement between data sets and an overall improvement of fit versus $\LCDM$. We expand our analyses to include a wide set of data, including the ACT-DR6 lensing data, as well as primary CMB information from ACT-DR4 and SPT-3G, scale-dependent power spectra from DES and measurements of $H_0$ from SH0ES. We find that there is little to no tension in measurements of structure within the data sets, and the inferred value of $S_8$ is generally lower than that in $\LCDM$. We find that the inclusion of DES generally favors a higher $H_0$, but there is some direct tension between the high-$\ell$ multipole data and this result. Future data should clarify whether this is a statistical artifact, or a true incompatibility of these datasets within this model.
\end{abstract}

\maketitle

\section{Introduction}
The era of precision cosmology has led to a remarkable quantitative understanding of the Universe. The experimental evidence for dark matter and dark energy has grown so strong that $\LCDM$ has been rightfully established as a Standard Model of cosmology. At the same time, the level of precision has allowed us to start looking for cracks in the cosmology, which may foretell physics beyond $\LCDM$.

Because of the expansive nature of these modern datasets, moving beyond $\LCDM$ often requires alternative scenarios. For instance, late universe modifications are often studied with wCDM or $w_0-w_a$ models. In the early universe, a number of models have been proposed, although there is not yet consensus on what alternatives might be most motivated, either theoretically or from data.

The motivation to understand such alternatives has been heightened by recent interest in ``tensions'' in $\LCDM$ \cite{snowmass}. Two tensions - the Hubble ($H_0$) tension, and the $S_8$ tension - in particular, have stood out and drawn attention as potentially foreshadowing new physics.

The Hubble tension is the significant difference between the value extracted from CMB data (in particular Planck \cite{planck}) in concert with results from other cosmological measurements such as Pantheon \cite{pantheon1, pantheon2} which aid in establishing the distance to the surface of last scattering, in comparison with the Cepheid-normalized value of $H_0$ from low-redshift supernovae \cite{sh0es,sh0es2}. 

The $S_8$ tension is the discrepancy between values of $S_8$ inferred within $\LCDM$ from measurements of the CMB \cite{planck, plancklens} and values extracted from weak lensing and galaxy counts \cite{DES, K1K}. For some time, this was viewed somewhat similarly to the Hubble tension, as a disagreement between early-universe observables in the CMB, and late-universe observables in large-scale structure. This has changed, however, with the arrival of the most recent results from ACT-DR6\cite{actdr6}. The measurement of CMB lensing, while using a light source originating at $z=1100$, is sensitive to late universe structure, and thus one cannot claim any longer there is a tension between late universe and early universe, but rather between LSS and other measurements. Indeed, in fits to $\LCDM$, the value of $S_8$, extracted from CMB lensing vs CMB power are quite compatible. 

Importantly, this is within $\LCDM$. The scales of fluctuations measured by CMB lensing and by LSS measurements of $S_8$ are somewhat different \cite{snowmass}, but are tightly related inside of $\LCDM$. In models beyond it, the tension may be ameliorated, or removed entirely.

As we have noted, many new models have been introduced to address cosmic tensions \cite{Brust_2017, Escudero_2020, Escudero_2020_2, Escudero_2021, Karwal_2016, Poulin_2019, Lin_2019, Smith_2020, Cyr_Racine_2022, Berghaus_2023, Niedermann_2020, Amon_2022, Cruz_2023, Buen-Abad_2022, Buen-Abad_2023}, and were further studied with the latest precision measurements \cite{nils_2021, khalife_2024, nils_2022, nils_2023, nanoom_2023, bagherian_2024, allali_2024, nils_2024}. Many of these models have implications for small scale structure as well, and can potentially address both tensions simultaneously. Indeed, it has been argued that the tensions are not entirely independent \cite{poulin_2024,bagherian_2024}. One such model, WZDR+ \cite{joseph}, builds on a scenario \cite{aloni} originally invoked to address the Hubble-tension. In it, there is a fluid of dark radiation to which dark matter is gently coupled. After the temperature drops below a mass threshold, that interaction turns off and the dark matter acts as cold dark matter. At scales inside the horizon before that, there is mild suppression of power. This model is an interesting alternative to consider irrespective of any tensions because it: i) is based on clear particle physics principles, ii) has a narrow set of new parameters, and iii) smoothly matches back onto LCDM within its parameter space.

In this paper, we revisit WZDR+ with the current suite of CMB data in order to find if the $S_8$ tension persists outside of $\LCDM$. We will study this scenario using current CMB data from Planck, ACT and SPT, along with Pantheon, and inferred power from DES. The layout of this paper is as follows: in Section \ref{sec:model} we outline the details of the WZDR+ model, including its particle content, interactions and effects on cosmology. In Section \ref{sec:analysis} we discuss our datasets in detail and show the implications of these datasets for WZDR+. We conclude our findings in Section \ref{sec:discussion} and discuss the outlook of cosmological tensions and search for new physics in upcoming experiments.

\section{A Dark Sector with Gentle Interactions}
\label{sec:model}
Typically, if dark matter scatters off itself or another fluid, we compare the scattering rate $\Gamma_{scat}$ to the Hubble rate. If $\Gamma_{scat} \ll H$ there is little change to the power spectrum of dark matter. On the other hand, if $\Gamma_{scat} \gg H$, the fluids become tightly coupled and dark acoustic oscillations result \cite{Cyr-Racine:2013fsa}. One can imagine that $\Gamma_{scat} \sim H$, but as $H\propto T^2$ is changing dramatically over much of cosmic history, it may seem at first unlikely that such a relation can be relevant for any significant period of time.

However, in many models, it is quite natural to have $\Gamma_{scat} \propto T^2 \propto H$. A model of dark matter scattering off a dark radiation bath via a massless vector or scalar gives precisely such a scaling. The relevance of such models cosmology were first studied by \cite{martin}. If such interactions persist until recombination, there can be tensions with LSS. However, as studied in \cite{joseph}, if the dark sector contains a mass threshold, the interaction can naturally turn off, gently suppressing power at scales that came inside the horizon prior to this transition, and yielding CDM on scales that came inside the horizon after the transition.

\subsection{Particle Content \& Interactions}
In this work, we will focus on the WZDR+ model, which is fully discussed in 
\cite{joseph}. From a particle physics perspective, the dynamics are governed by the initial conditions (the amount of initial dark radiation), the dark interactions (parameterized by various coupling constants), and the mass scales of the dark sector (of which we take only $m_\phi$ to be relevant). These map onto the more physical relevant parameters $\NIR$ (the amount of dark radiation at late times), $z_t$ (the redshift when dark interactions turn off), and $\Gamma_0$ (a late universe parameterization of the scattering rate between dark matter and dark radiation ).

A concrete realization consists of a dark sector containing a complex scalar $\phi$ with mass $m_\phi$, as well as a massless Weyl spinor $\psi$. As stated in \cite{aloni}, a Yukawa coupling of the form $\LL \supset \lambda_\psi \phi \psi^2$ can be generated in a Wess-Zumino supersymmetric model, where the masses of the two species also arise quite naturally. The dark sector couples to dark matter particles $\chi$, here assumed to be fermions with mass $M_{\chi}$, via an additional Yukawa coupling $\LL \supset \lambda_{\chi}\phi\chi^2$. Depending on the coupling, energy and momentum transfer between dark matter and dark radiation can be significant through t-channel scattering ($\psi\chi\longrightarrow\psi\chi$), where $\phi$ particles from the dark radiation serve as virtual mediators.

Much like the presence of many mass threshold within the Standard Model, the mass threshold $m_\phi$ within WZDR+ generates rich dynamics both within the dark radiation and with dark matter, and is thus worth reviewing here. In this model, we assume an initially self-thermalized dark sector at early times with temperature $\TDS \gg m_{\phi}$, which may be less than the photon temperature $T_\gamma$.  In this limit, both $\psi$ and $\phi$ are relativistic and can maintain chemical ($\phi\longleftrightarrow\psi\psi$, $\phi\phi\longleftrightarrow\psi\psi$) and kinetic ($\phi\psi\longrightarrow\phi\psi$) equilibria. It is worth noting that both CMB and big bang nucleosynthesis (BBN) place stringent bounds on the early time value of $\Neff$ beyond the Standard Model value, or equivalently the amount of additional thermalized radiation at early times. However, in many models it is natural for the dark sector to equilibrate with the SM only after BBN. For instance,  this scenario can be generated across a wide parameter space of a dark fermion-neutrino oscillation model \cite{BBNStep}. So long as this equilibration happens below $\sim \MeV$, BBN bounds are easily avoided \cite{Giovanetti:2024orj}.

The dark sector temperature redshifts with the scale factor $a$ as $\TDS \propto a^{-1}$ for most of its history. However, in the vicinity of $\TDS \sim m_{\phi}$, the scalar $\phi$ becomes non-relativistic, and subsequent falls out of chemical equilibrium with $\psi$ at some temperature $\TDS \lesssim m_{\phi}$. Since the sector is thermalized, the comoving entropy density must be conserved, and as a consequence the quantity $a \TDS$ goes through a ``step" of the following form
\begin{align} \label{eq:TDS}
    (a\TDS)^{\rm UV} &= \left(\frac{g_{*s}^{\text IR}}{g_{*s}^{\text UV}}\right)^{1/3}(a\TDS)^{\rm IR}\\
    &\equiv (1+r_g)^{-1/3}(a\TDS)^{\rm IR}\, , 
    \nonumber
\end{align}
where here UV and IR refer to the temperature above and below the mass threshold respectively, and $r_g = g^{\phi}_* / g^{\psi}_*$ is the  ratio of the relativistic degrees of freedom of the dark species ($r_g=8/7$ for WZDR+). One sees then that the scalar $\phi$ annihilation injects its energy into the remaining $\psi$ fluid, heating the dark sector. This process is completely analogous to the heating of Standard Model photons at a threshold photon temperature $T_{\gamma} = m_e$, due to $e^+ e^-$ annihilation. The dark radiation self interaction is assumed to be strong enough such that the freeze-out relic abundance of $\phi$ is negligible, and the remaining $\psi$ particles continue to strongly self interact via an effective 4-Fermi operator.

For convenience, we parameterize the dark radiation energy density in terms of the effective number of addition neutrino species $N_{\rm UV/IR}$ before and after the threshold. Similar to Eq.~\ref{eq:TDS}, they are related by
\begin{align} \label{eq:Nstep}
    \frac{N_{\rm IR}}{N_{\rm UV}} = (1+r_g)^{1/3}\, ,
\end{align}
resulting in a ``step" in the dark sector population (i.e., a change in $\Neff$). Given this mass threshold generated dynamics, two of the three parameters in the WZDR+ are 1) $m_\phi$, which controls the temperature where the threshold transition occurs and 2) $N_{\rm IR}$, which fixes the total amount of dark radiation at late times (as well as $N_{\rm UV}$ at early times, as given by Eq.~\ref{eq:Nstep}). We follow the convention of past analyses of WZDR+ and recast the $m_\phi$ parameter as a transition redshift $z_t$ (see Appendix. C of \cite{aloni}). We note that \cite{joseph, aloni} have studied an extended model where the relative degrees of freedom between $\phi$ and $\psi$ is a variable, in which case both $N_{\rm UV}$ and $N_{\rm IR}$ are free parameters. However, the past studies did not find any preference for the extra degree of freedom over the original WZDR, and thus we do not consider this extension in this work. 

Finally, the effect of the WZDR+ step is also significant on the dark matter-dark radiation scattering. Since dark matter couples to the massless fermions via the massive $\phi$, the Hubble-normalized scattering rate $\Gamma/H$ drops precipitously for $\TDS < m_\phi$. The most efficient contribution to dark matter momentum transfer is $\psi\chi\longrightarrow\psi\chi$ process mediated by a t-channel $\phi$. The similar s-channel process $\phi\chi\longrightarrow\phi\chi$ is subdominant due to suppression by the dark matter mass in the propagator. Interestingly, \cite{martin} has shown that the dark matter momentum transfer rate via the t-channel process is proportional to $\TDS^2/M_\chi$. This means that at early times, the ratio of the momentum transfer rate to the Hubble rate scales as
\begin{align}\label{eq:GammaOverH_early}
    \frac{\Gamma}{H} \propto  \left(\frac{\TDS}{T_\gamma}\right)^2\, ,
\end{align}
where $T_\gamma$ is the Standard Model bath temperature. Away from any Standard Model or dark sector mass thresholds, both temperatures redshift, meaning $\Gamma/H$ remains constant at early times. This dark matter scattering imparts an effective pressure within the dark matter fluid, which could have significant effects on the growth of structures. To date, the rich ensemble of large-scale structures observations place a stringent upper bound on the interaction strength of dark matter, beyond which certain structures would not have formed. The WZDR+ model naturally reconciles this observable by rapidly shutting off the dark matter interaction after the step. For $\TDS < m_\phi$, the dark matter momentum transfer rate now suffers by an additional $(\TDS/m_\phi)^4$ suppression, coming from a dimension-6 $\psi^2 \chi^2$ 4-Fermi operator, and thus quickly becomes irrelevant. 

\subsection{Effect on Cosmology}
\begin{figure*}[t]
    \centering
    \includegraphics[width=0.6\linewidth]{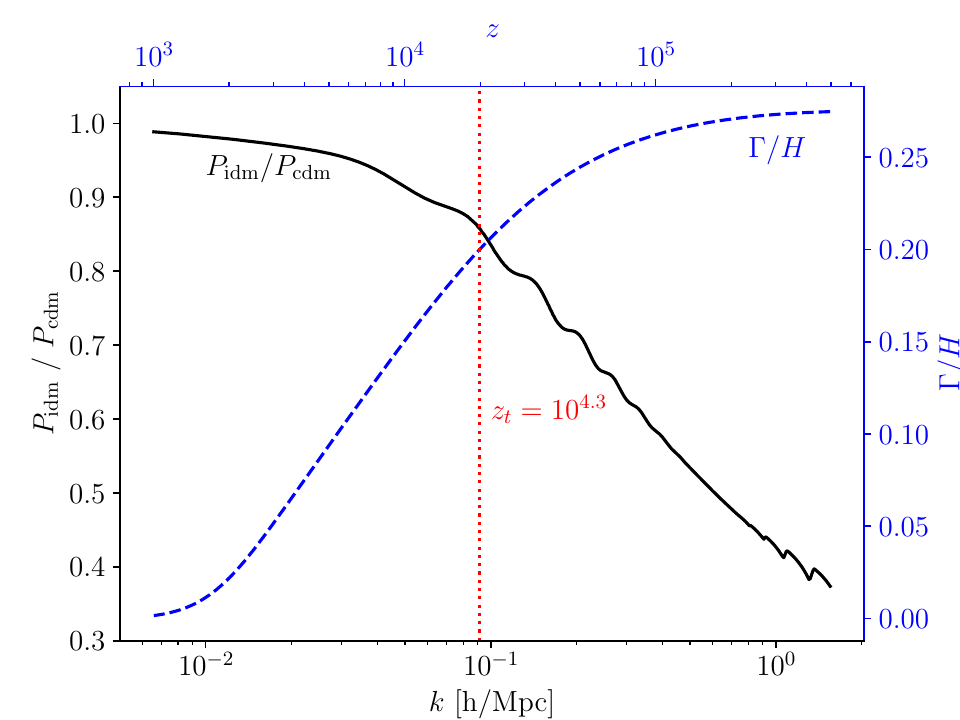}
    \caption{Visualization of the dark matter scattering effect on cosmology. The solid black curve is given by the ratio of the WZDR+ linear MPS ($P_{\rm idm}$) to that of $\LCDM$ ($P_{\rm cdm}$) evaluated at $z=0$, with both models assuming the same background cosmology. 
    The dashed blue curve shows the ratio of the WZDR+ dark matter momentum transfer rate $\Gamma$ to Hubble over redshifts $z$. The fourier scales $k$ on the bottom axis are aligned with redshifts on the top axis, such that each value of redshift $z$ is the horizon crossing time of the scale $k$ on the bottom. The dotted red line marks the transition redshift $z_t=10^{4.3}$ used to compute this figure. Small scales that enter the horizon while the scattering rate is significant suffer a suppression of power, while large scales that grow after the transition behave effectively like cold dark matter. In this figure we assumed $\omega_b=0.02272$, $\omega_{\rm dm}=0.1288$, $h=0.7135$, $A_s=2.1\times 10^{-9}$, $n_s=0.9660$, $\tau_{\rm reio}=0.0586$, $z_t=10^{4.3}$, $N_{\rm IR}=0.6$, $\Gamma_0=10^{-6}\ {\rm Mpc}^{-1}$, $Y_p=0.2455$, and one massive neutrino at $m_\nu=0.06\ {\rm eV}$ such that the standard model $N_{\rm eff}=3.044$.}
    \label{fig:rate}
\end{figure*}

Obtaining the momentum transfer rate at arbitrary redshift requires solving a difficult Boltzmann equation. A convenient approximation of the result can be written as 
\begin{align}
    \Gamma(x) = \Gamma_0  \frac{(1+z_t)^2}{x^2}\left(\frac{1}{1-0.05\sqrt{x} + 0.131x}\right)^4\, ,
    \label{eq:DMDRrate}
\end{align}
where $x\equiv m_\phi/\TDS$, $z_t$ is the transition redshift when $x=1$, and $\Gamma_0$ (in units of [${\rm Mpc}^{-1}$]) is the third free parameter of the WZDR+ that sets a reference transfer rate. Fig.~\ref{fig:rate} shows the momentum transfer rate evolution compare to the Hubble rate, and the corresponding effect on the linear matter power spectrum (MPS). Referenced to a $\LCDM$ predicted MPS, the MPS in WZDR+ suffers a suppression of power at small scales. This can be understood as the resistance to clustering at high temperatures due to the presence of ``gentle" dark matter scattering, at an early time where the size of the horizon corresponds to these small scales. In the opposite limit, WZDR+ is able to recover a $\LCDM$-like MPS at large scales, as the dark matter interaction has long been shut off when these modes have entered the horizon at late times. 

Due to this feature, WZDR+ is particularly interesting for the $S_8$ tension. It is worth noting that early universe surveys of clustering, either via CMB anisotropies, or through LSS surveys such as BOSS \cite{Alam_2017}, do not directly measure the amplitude of the MPS at scales as small as $k = (8\ {\rm Mpc}/h)^{-1}$. Rather, LSS observables are most sensitive to the power at generally larger scales, and $S_8$ is subsequently extrapolated in a model dependent procedure, where typically a $\LCDM$ cosmology is assumed. Thus, a MPS which behaves according to Fig.~\ref{fig:rate} is significant for deducing $S_8$. While both $\LCDM$ and WZDR+ will calibrate the amplitude of MPS to similar values according to the large scale set by LSS observations, WZDR+ will generically extrapolate a smaller $S_8$ value compared to $\LCDM$.

Lastly we outline the relevant cosmological perturbations. For the dark radiation itself, we follow the conventional treatment of a strongly self interacting relativistic species, for which only the density perturbation $\delta_{\rm dr}$ and velocity perturbation $\theta_{\rm dr}$ are important. For the interacting dark matter, we track its density perturbation $\delta_{\rm idm}$ and velocity perturbation $\theta_{\rm idm}$, and assume negligible dark matter sound speed. We use adiabatic initial conditions for all perturbations. Working entirely in the synchronous gauge \cite{Ma_1995}, these perturbations evolve according to the following set of four coupled differential equations
\begin{align}\label{eq:perturbations}
    \nonumber \Dot{\delta}_{\rm dr} &= -(1+w)\left(\theta_{\rm dr} + \frac{\Dot{h}}{2}\right) - 3\HH(c_s^2 - w)\delta\, ,\\
    \nonumber \Dot{\theta}_{\rm dr} &= \frac{k^2 c_s^2}{1+w}\delta_{\rm dr} - \HH(1-3 c_s^2)\theta_{\rm dr} + a\Gamma R(\theta_{\rm idm} - \theta_{\rm dr})\, ,\\
    \nonumber \Dot{\delta}_{\rm idm} &= -\left(\theta_{\rm idm} + \frac{\Dot{h}}{2}\right)\, ,\\
    \Dot{\theta}_{\rm idm} &= -\HH \theta_{\rm idm} + a\Gamma(\theta_{\rm dr} - \theta_{\rm idm})\, ,
\end{align}
where $\ \Dot{}\ $ denotes derivative with respect to conformal time, $\HH = a H$ is the conformal Hubble parameter, $h$ is the metric perturbation\footnote{$h$ is the metric perturbation in this equation only. For every other occurrence of $h$ in this work, it is the dimensionless Hubble parameter.} in the synchronous gauge, $w$ and $c_s$ are the dark radiation equation of state and sound speed respectively, $\Gamma$ is the momentum transfer rate at current time given by Eq.~\ref{eq:DMDRrate}, and $R\equiv \rho_{\rm idm}/(\rho_{\rm dr} + P_{\rm dr})$.

\section{Analysis}
\label{sec:analysis}

\subsection{Likelihoods \& Methods}
\label{sec:methods}

We compute cosmological perturbations of WZDR+ with a modified version of CLASS v3.1 \cite{CLASS}, including Eqs.~\ref{eq:DMDRrate} \& \ref{eq:perturbations} numerically. We perform Markov-chain Monte Carlo (MCMC) sampling using the MontePython v3.5 \cite{montepython1, montepython2} package with a Metropolis-Hastings algorithm. The WZDR+ free parameters we sample  are the six $\LCDM$ free parameters  $\{\omega_b, \omega_{\rm dm}, 100\theta_s, \ln{(10^{10} A_s)}, n_s, \tau_{\rm reio}\}$\footnote{$\omega_{\rm dm}$ refers to cold dark matter for $\LCDM$ and interacting dark matter for WZDR+.}, as well as three extra parameters $\{\log_{10}{(z_t)}, N_{\rm IR}, \Gamma_0\}$, all with uniform priors. We assume 100\% interacting dark matter in WZDR+ (set $f_{\rm idm}=1$ in CLASS). In our fits we fix the primordial helium abundance $Y_p=0.245$ rather than the standard BBN computation by CLASS, as to simulate a dark sector that thermalizes after BBN. We follow the convention of the Planck analysis \cite{planck} and include one massive neutrino with $m_\nu = 0.06$ eV and two massless neutrinos, ensuring that the total standard model $N_{\rm eff}=3.044$. We use HALOFIT \cite{halofit1, halofit2} to compute non-linear corrections for the MPS. 

The list of datasets we consider are as follows. Our baseline dataset $\boldsymbol{\DD}$ includes
\begin{itemize}
    \item \textbf{Planck CMB Anisotropy} \cite{planck}, including both high-$\ell$ TT, TE, EE and low-$\ell$ TT, EE, with the full set of nuisance parameters.
    \item \textbf{CMB Lensing}. We use a joint likelihood that includes both Planck 2018 and ACT-DR6\footnote{We used the numerical likelihood available at \url{https://github.com/ACTCollaboration/act_dr6_lenslike}} lensing results \cite{plancklens, actdr6}. 
    \item \textbf{BAO} with the BOSS-DR12 \cite{Alam_2017} BAO-only likelihood(z=0.38, 0.51, 0.61), as well as the BAO small-z 2014 likelihood, which includes 6dF \cite{6DF} (z = 0.106) and MGS \cite{MGS} (z = 0.15) catalogues.
    \item \textbf{PANTHEON} supernova calibration likelihood \cite{pantheon1}.
\end{itemize}
We note that our dataset $\boldsymbol{\DD}$ differ from those of \cite{joseph, aloni}, \cite{nils} by the addition of ACT-DR6 lensing likelihood. For selective tests, we also include the following additional measurements of the CMB temperature and polarization anisotropies
\begin{itemize}
    \item \textbf{ACT} from ACTPol-DR4 \cite{actdr4}, with $\ell$-modes overlapping with Planck 2018 removed. We note that the dataset name ``ACT" in the rest of this work refer solely to this primary CMB measurement, and not to the aforementioned lensing measurement from ACT-DR6.
    \item \textbf{SPT} from the SPT-3G Y1 data release \cite{SPT3G}.
\end{itemize}

Our list of late universe measurements of $H_0$ and $S_8$ include

\begin{itemize}
    \item {$\boldsymbol{\HH}$} Hubble prior measured by the SH0ES collaboration \cite{sh0es} with value of $h=0.7304\pm 0.0104$.
    \item {$\boldsymbol{{\rm DES}}$} Matter power spectrum from DES Y3 \cite{Doux_2022}, reconstructed cosmic shear measurements. See Sec.~\ref{sec:des} for a more detailed discussion.
\end{itemize}

For fitting models to each dataset, we run 8 parallel MCMC chains, and consider chains converged by using the Gelman-Rubin convergence test, $R-1 < 0.01$, for all parameters. For finding the minimum best-fit points and their $\chi^2$ values, we use an optimizer made from modifying the Metropolis-Hastings algorithm to only accept proposed steps with a lower $\chi^2$, removing the random aspect of the sampler. Note that this modification is used only after we fully explore the parameter space with the unmodified Metropolis-Hastings sampler. 

It is also worth noting here that MontePython automatically compute the parameter credible intervals from the posterior histograms. By default, the algorithm locates the highest density intervals that contain 68\%, 95\% and 99.5\% of the integrated probability density function, and reports the bounds of these intervals as the $\sigma$-levels. In this work we encounter numerous non-Gaussian posterior distributions where this method is not appropriate. This is because the highest density region is dependent on the actual height of the distribution at particular points, which is not invariant under reparameterization. For the rest of this work, our reported $\sigma$-levels are obtained based on the area excluded within the cumulant distribution function, the reparameterization invariant quantity. For instance, our 1-$\sigma$ points are such that 16\% of the area under the distribution are excluded to either side, leaving 68\% credibility within the 1-$\sigma$ interval. In cases where the distribution is one sided without a peak, we instead report the 95\% upper bound.

\subsection{Treatment of the $S_8$ Likelihood}\label{sec:des}
Conventionally, studies of beyond $\LCDM$ models that address the $S_8$ tension have directly quoted the $S_8$ values reported by late universe surveys. Notable results include $S_8=0.766^{+0.020}_{-0.014}$ from the KiDS-1000 collaboration \cite{K1K}, and $S_8=0.775^{+0.026}_{-0.024}$ from DES Y3 \cite{DES}. These are often implemented as asymmetric Gaussian priors for the MCMC sampling, and used to compare against the corresponding posteriors. In this section, we address the shortcomings of this approach and the alternative we consider in this work.

It is important to understand that cosmic shear experiments such as KiDS-1000 and DES do not explicitly measure the quantity $S_8$. Instead, they measure the shear angular power spectrum over some range of small angular scales, which do not directly correspond to the Fourier scale of $k=8 {\rm Mpc}^{-1}$ used to define $S_8$. The exercise of extracting $S_8$ from shear measurement therefore requires one to assume a cosmological model, and rely on its transfer function to infer the value of the matter power spectrum at a particular scale. The KiDS-1000 analysis for instance adopts the ``3x2pt" model and performs a nested sampling fit to their shear measurements, with 5 $\LCDM$ cosmological parameters and 15 nuisance parameters unique to their likelihoods. The resulting reported value of $S_8$, therefore, is an inferred value assuming $\LCDM$ to be the background cosmology, and in principle should not be used to inform or to make statements about non-$\LCDM$ models with modified transfer functions. We further point out that the ongoing $S_8$ tension is not an intrinsic tension between early and late universe measurements, but instead a model dependent tension between the extrapolated values of $S_8$ assuming $\LCDM$ when fitted to these measurements.

For these reasons, we do not quote the existing $S_8$ values in literature when testing WZDR+. We also do not implement these values as likelihoods used in fits. In place of an $S_8$ prior, we use the DES Y3 measurements in the form of binned linear MPS data points, which are reconstructed from the angular shear power spectra\footnote{See Fig. 17 of \cite{Doux_2022}. The data underlying this likelihood were obtained through private communications with members of the DES collaboration. These data can be provided by the DES collaboration upon request.}. As pointed out by \cite{Doux_2022}, the $C_{\ell}\rightarrow P(k)$ reconstruction is itself a model dependent process. In particular, the mapping of $k=(\ell+\frac{1}{2})/\chi(z)$ is sensitive to the calibration of the angular diameter distance in the background cosmology. However, this distance is mostly calibrated by late time measurements and left unmodified by the early time dynamics in WZDR+. The extrapolated values are therefore appropriate to use here, although less optimal than a dedicated fit using the shear $C_{\ell}$'s directly. This is similar to the analysis of  \cite{bagherian_2024}, who used the amplitude and slope of the power in Ly-$\alpha$ data to study WZDR+.

We implement a multivariate Gaussian likelihood using the five binned data points with the lowest variance, which are the central bins ranging from $0.0178\ {\rm h/Mpc} \leq k \leq 1.78\ {\rm h/Mpc}$. We refer to this $P(k)$ likelihood as simply ``DES" for the rest of this work. 

\subsection{Results}

As discussed in section \ref{sec:methods}, we separate our different datasets into simple categories, to which we here add some additional perspective.
 We choose the dataset $\DD$ to be a baseline representation the early universe data, although we have included within $\DD$ the lensing likelihoods from ACT-DR6, as it has been in agreement with Planck within $\LCDM$. This combines a set of well-understood datasets with extensive study.
 We also have early universe data from ACT-DR4 and SPT-3G, but these datasets are not as vetted and understood as others. There are no reasons to doubt these data, but in light of this and tensions with other data, it is helpful to separate them out. Given the well known Hubble tension, we also leave out the $H_0$ prior from SH0ES for the moment.

\begin{figure}[h]
    \centering
    \includegraphics[width=\linewidth]{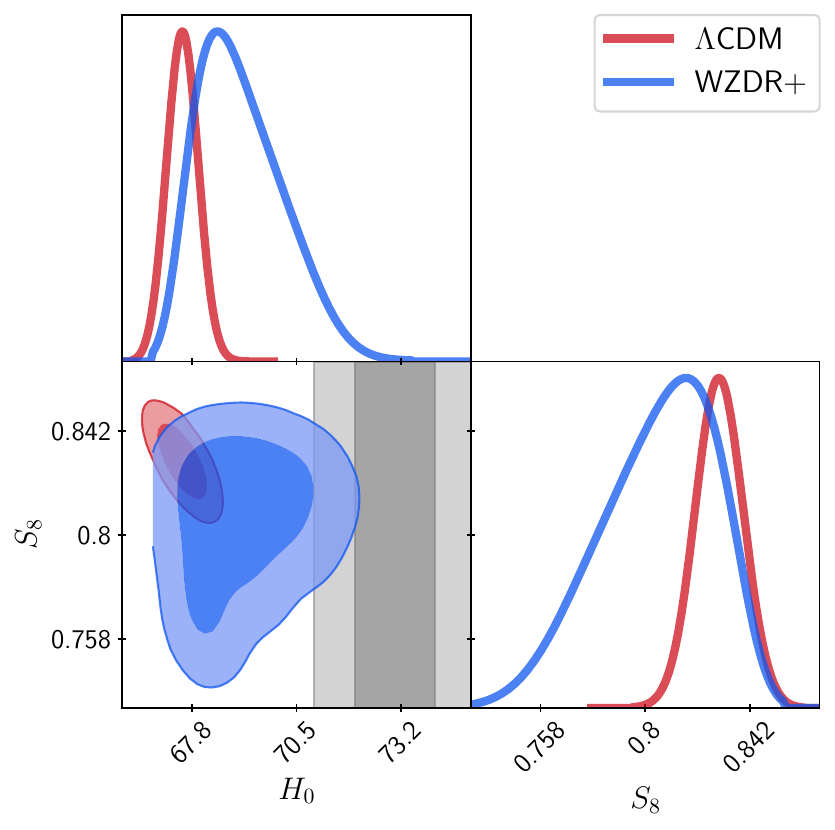}
    \caption{Posterior in the $H_0$-$S_8$ plane, for $\LCDM$ and WZDR+ fitted to dataset $\DD$. The grey band shows the $1\sigma$ and $2\sigma$ regions of the SH0ES measurement.}
    \label{fig:LCDMvWZDR+_H0S8}
\end{figure}

 Within this baseline $\DD$ dataset we find 
\begin{align}
    \Omega_m &= 0.3088_{-0.006}^{+0.0061}, \hskip 0.1in
    H_0 = 69.07_{-1.1}^{+1.2},  \nonumber \\
    S_8 &= 0.8056_{-0.025}^{+0.023}, \hskip 0.2in[\DD]\nonumber
\end{align}
where we have reported the mean and 1-$\sigma$ deviations for the derived parameters $\Omega_m$, $\sigma_8$ and $S_8$. We additionally perform a $\LCDM$ fit to the same dataset, and in Fig.~\ref{fig:LCDMvWZDR+_H0S8} we compare the posterior distributions of the two models in the $H_0$ and $S_8$ parameter plane. We see the $S_8$ distribution for WZDR+ is significantly wider than that for $\LCDM$, with the peak shifted to a lower value relative to the $\DD$ result for $\LCDM$ of $S_8=0.8297^{+0.0097}_{-0.0096}$.

In some sense, this is one of our primary results. Even before consulting additional datasets we see that the value of $S_8$ in this model has both a lower mean and, more importantly, a significantly wider range of allowed values. This suggests that the tensions between extracted values of $S_8$ from $\DD$ in comparison with weak lensing and galaxy count data are at the very least model dependent. In particular, the CMB lensing results of ACT-DR6 are still compatible with a wide range of values of $S_8$. 

WZDR+ combined $\DD$ also has a wider allowed range of values of $H_0$, which is consistent with previous results\cite{joseph} where ACT-DR6 was not included. 

The overall fit of WZDR+ is not significantly better ($\Delta \chi^2_{\rm tot} \sim -1$ for three new degrees of freedom), so these datasets alone do not have any preference for new physics. 

This wider range of $S_8$ and $H_0$ is suggestive that WZDR+ provides a better fit to the overall data than \lcdm. We can do this by including additional data on galaxy structure and/or data from SH0ES.

We will begin by incorporating $S_8$ measurements. As we have noted, there is no direct measurement of $S_8$ from DES or KiDS within the WZDR+ model. However, we can combine $\DD$ with 
a late universe MPS likelihood from DES Y3. This combination  yields
\begin{align*}
    \Omega_m &= 0.303_{-0.0061}^{+0.006},\hskip 0.1in 
    H_0 = 70.9_{-1.4}^{+1.4},\\
    S_8&= 0.7872_{-0.018}^{+0.017}.\hskip 0.2in  [\DD\text{+DES}]
\end{align*}
The values of both $\Omega_m$ and $S_8$ are compatible at better than 1$\sigma$. The central value shifts down, as expected. 

A concrete way to quantify WZDR+'s performance against $\LCDM$ is through locating their best-fit points for a given data set, and using these points to compute the $\Delta$AIC (Akaike information criterion, see \cite{nils_2021} for a discussion). It is defined as 
\begin{align}
    \Delta {\rm AIC} = \chi^2_{\rm WZDR+} -\chi^2_{\LCDM} + 2\times N_{\rm extra}\, ,
\end{align}
where $N_{\rm extra}=3$ is the number of extra parameters in WZDR+ compared to $\LCDM$. Thus, lower scores denote an overall better fit to the data than $\LCDM$, which is only achievable for $\Delta \chi^2$ reached beyond the penalty for having extra parameters. We find $\Delta\rm{AIC}=+5.26$ for $\DD$. This positive and order unity scores suggest that WZDR+ fits to early universe data at a comparable level as $\LCDM$. However, the scores lowers considerably to $\Delta\rm{AIC}=-18.84$ once we add in the DES data, signifying that WZDR+ is able to accommodate the combination of early and late universe measurements much better than $\LCDM$. We provide a comprehensive summary of best fit points, $\chi^2$ and $\Delta$AIC in Table.~\ref{tab:bf}.

$H_0$ is an interesting story. As previously found for the dataset $\DD$, there is a broader uncertainty within WZDR+ ($\sim 1.2\ {\rm vs}\ 0.4$) and a higher best fit (68.64 vs 67.56) relative to \lcdm. This is in better agreement with SH0ES than \lcdm. However, the best fit value of $\nir=0.154$ is still within 1$\sigma=0.16$ of 0.

The addition of the DES data seems to change the situation. The best fit value of $H_0$ in $\DD$+DES shifts to 71.10, a roughly $1.7 \sigma$ shift. The $1\sigma$ range in $H_0$ easily overlaps the preferred range from SH0ES. There is some mild preference for new physics in this dataset, evidenced by the best fit value of $\nir=0.569$, which is approximately $2.3\sigma$ from 0. The best fit value of $10^6 \Gamma_0 = 0.586$ is similarly almost $2\sigma$ from 0.

If the story ended here, it might appear there was a concrete direction of the data - namely, that within WZDR+, there was a consistent indication for additional interacting radiation, and a gentle interaction with dark matter, turning off at a redshift near $z=20,000$. However, the inclusion of additional primary CMB data from ACT and SPT seem at odds with this interpretation, specifically with regards to $H_0$.

We can step back and include primary CMB measurements from ACT and SPT to our baseline dataset $\DD$. We find that fitting WZDR+ to $\DD$+ACT+SPT yields 
\begin{align}
    \Omega_m &= 0.3114_{-0.0056}^{+0.0056}, \hskip 0.1in 
    H_0 = 68.19_{-0.67}^{+0.68}, \nonumber \\
    S_8&=0.7965_{-0.027}^{+0.025}.\hskip 0.1in  [\DD\text{+ACT+SPT}] \nonumber
\end{align}

\begin{figure*}[t]
    \centering
    \includegraphics[width=0.6\linewidth]{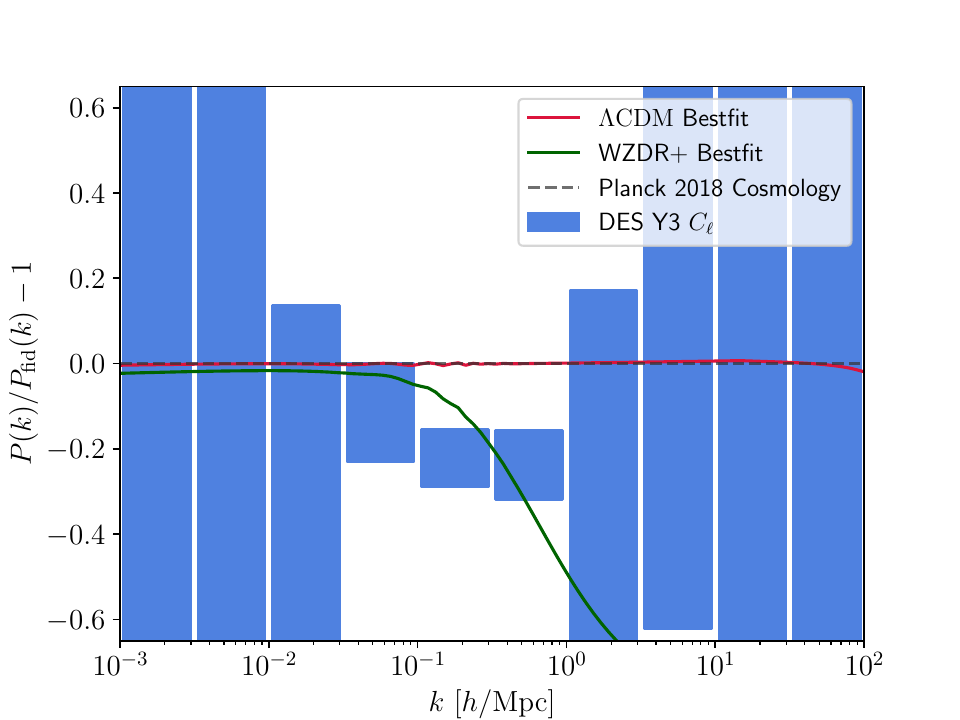}
    \caption{Matter power spectra from the $\LCDM$ and WZDR+ best fits to $\DD$+ACT+SPT, compared to DES Y3 measurements. The power spectra are normalized to a fiducial matter power spectrum $P_{\rm fid}(k)$ used in the DES analysis, obtained from a Planck 2018 fiducial cosmology. The black dashed line at $y=0$ denotes this benchmark. The red and green curves denote the $\LCDM$ and WZDR+ best fits respectively. The blue boxes are the DES data points inferred from angular shear $C_{\ell}$'s, where the width denote the log-$k$ bin widths, and the heights denote the uncertainty.}
    \label{fig:desredofig}
\end{figure*}

Again, we find a lower value of $S_8$ with a broader uncertainty. 
The addition of ACT and SPT shift the $S_8$ value lower compared to $\DD$, by about $0.5\sigma$. Indeed, the fit of $\DD$+ACT+SPT provides a good fit to DES data, as one sees in Fig.~\ref{fig:desredofig}.

$H_0$, however, shifts to the lower value 68.19 with a narrower spread of $\sigma \sim 0.68$. The best fit value of $\nir$ is now shifted to lower values, with a best fit of $0.012$, which is essentially indistinguishable from 0. While the error bars are larger than \lcdm, the overall push of these additional CMB data is to reinforce the Hubble tension. 

We can now consider the full set of data in our analysis -  $\DD$+ACT+SPT+DES - and find
\begin{flalign*}
    \Omega_m &= 0.3073_{-0.0055}^{+0.0054}, \hskip 0.1in 
    H_0 = 69.2_{-1.2}^{+1.2}, \nonumber \\
    S_8&= 0.7778_{-0.018}^{+0.017}. \hskip 0.1in [\DD\text{+ACT+SPT+DES}] \nonumber
\end{flalign*}
The $S_8$ value obtained between all these different combinations remains largely consistent. The $\Delta$AIC=-16.58 is still significant, indicating this model provides a much better fit to the cumulative data than \lcdm\, alone.  

\begin{figure}[H]
    \centering
    \includegraphics[width=\linewidth]{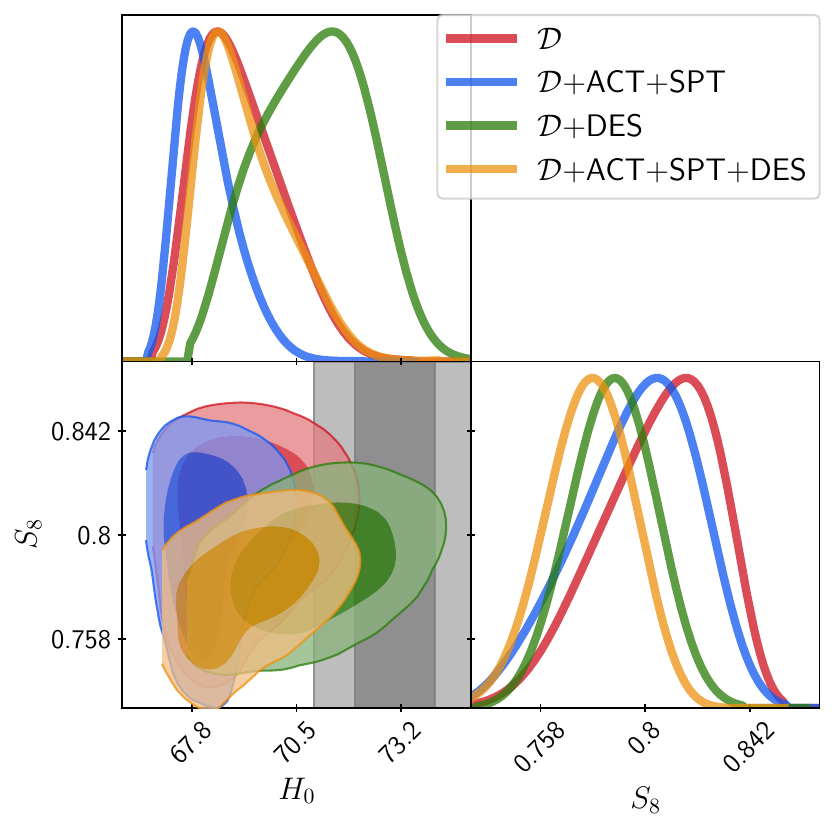}
    \caption{Posterior in the $H_0$-$S_8$ plane, for WZDR+ fitted to four dataset combinations. For each dataset, the dark and light contours correspond to the 68\% and 95\% confidence regions, respectively. The grey band shows the one and two sigma ranges from the SH0ES direct measurement.}
    \label{fig:D_s8}
\end{figure}

In contrast, the DES and SPT+ACT data appear to pull $H_0$ in opposite directions, yielding a distribution largely unchanged from the initial $\DD$ data, alone. 
 We show a comparison of the four fits in Fig.~\ref{fig:D_s8}.

\section{Discussion}
\label{sec:discussion}
The era of precision cosmology is yielding a new period where models beyond $\LCDM$ can be studied, and quantitative comparisons made. Although there are many parametric extensions to $\LCDM$, there is a narrower set that is grounded in well-understood particle physics, and which provide superior fits to some data sets. In this paper, we have considered one such model, WZDR+. With the presence of gentle interactions between dark matter and dark radiation, coupled with a mass threshold which turns off those interactions, WZDR+ naturally recovers the long-distance successes of $\LCDM$ while providing non-trivial modifications at early times and short distances.

An important consequence of this is that the tensions between different scales of structure present in $\LCDM$ are absent in WZDR+. Even the newest CMB lensing results from ACT-DR6 are compatible with a much broader range of $S_8$. We have incorporated a scale-dependent power spectrum from DES as well and find no tension between the data sets. Moreover, we find WZDR+ improves the fit to these data dramatically, improving $\chi^2$ by 22, yielding  $\Delta{\rm AIC}=-16.58$. 

\begin{figure}[h!]
    \centering
    \includegraphics[width=\linewidth]{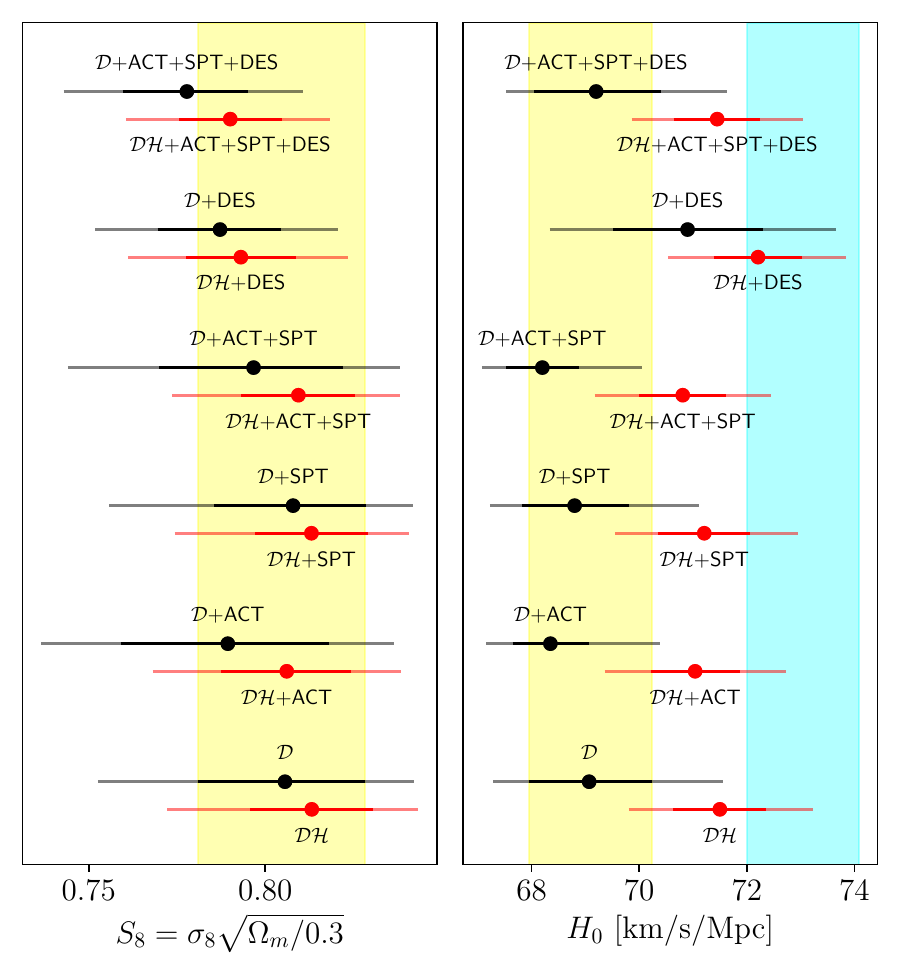}
    \caption{Summary figure with mean$\pm\sigma$ intervals of $S_8$ and $H_0$, for WZDR+ fitted to a combination of CMB+BAO, cosmic shear, and the hubble measurement from SH0ES. Solid and faded error bars at each point represent the 1 and 2-$\sigma$ credible intervals respectively. The black error bars denote fits without the SH0ES prior, and the accompanying red error bars are with the prior added. The background yellow bands are 1-$\sigma$ extensions of our fit to the default dataset $\DD$. In the right panel for $H_0$, the cyan band denotes the direct Hubble measurement from SH0ES.}
    \label{fig:whisker}
\end{figure}

Importantly, we do {\it not} study the overall fit of the model by including an $S_8$ posterior derived in $\LCDM$. While many analyses have previously used such priors, one should recognize the significant caveats that accompany this. Because the amplitudes of fluctuations at different scales are tightly correlated within $\LCDM$, one can yield a narrow value for $S_8$, even if the actual scales probed have only minimal overlap with the window function for $S_8$. In alternative models, this tight relationship is broken, and the different datasets are easily in accord. We show the various extracted values of $S_8$ from the different combinations of data in Fig. \ref{fig:whisker}.

\begin{figure}[h!]
    \centering
    \includegraphics[width=\linewidth]{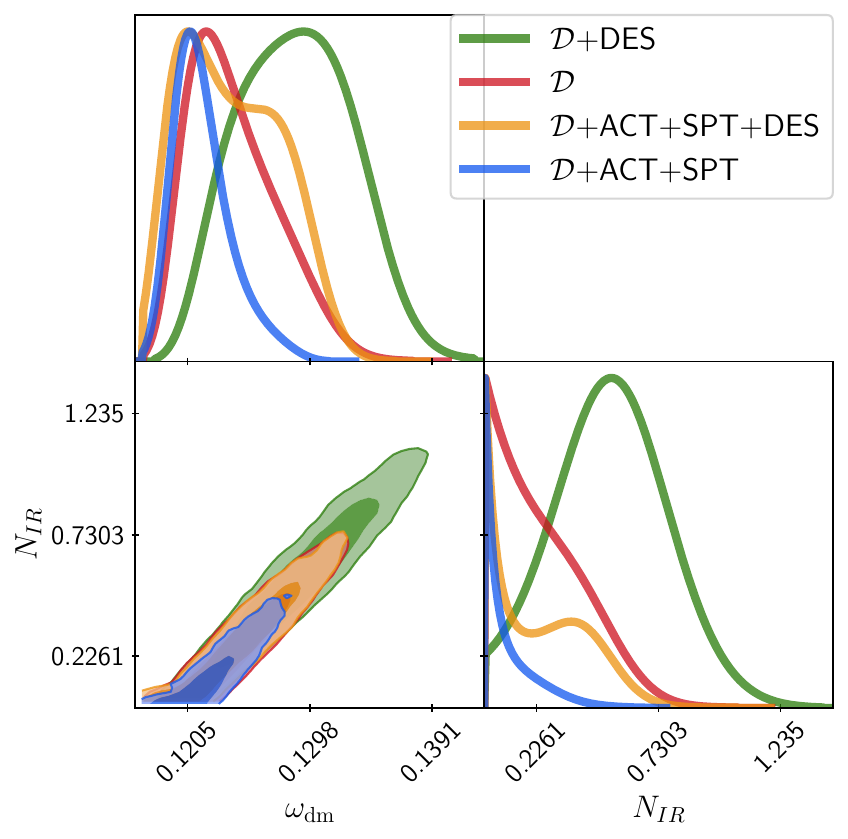}
    \caption{Posterior distributions of $\omega_{\rm dm}$ and $N_{\rm IR}$ for WZDR+ fitted to 4 datasets. We note that the multimodal behavior of the $\DD$+ACT+SPT+DES dataset is due to internal conflicts between ACT+SPT vs. DES.} 
    \label{fig:dm_v_nir}
\end{figure}

The Hubble tension is a more complicated question, however. While WZDR+ has a broader range of allowed values of $H_0$ compared with \lcdm, it cannot be claimed to directly ``solve'' the tension, but only ameliorate it. While the DES data seem to favor a higher value of $H_0$, the ACT and SPT data, taken together, favor a lower value. The combination of these along with $\DD$ leave the overall posterior for $H_0$ largely unchanged from $\DD$, alone. Related to this conflict in $H_0$, we find that ACT and SPT prefer lower values for both $\omega_{\rm dm}$ and $N_{\rm IR}$, while DES favors the opposite. This is shown in Fig.~\ref{fig:dm_v_nir}. These datasets end up pushing the posteriors in opposite directions with comparable statistical weights, resulting in multimodal distributions in the joint fit.

\begin{figure}[h!]
    \centering
    \includegraphics[width=\linewidth]{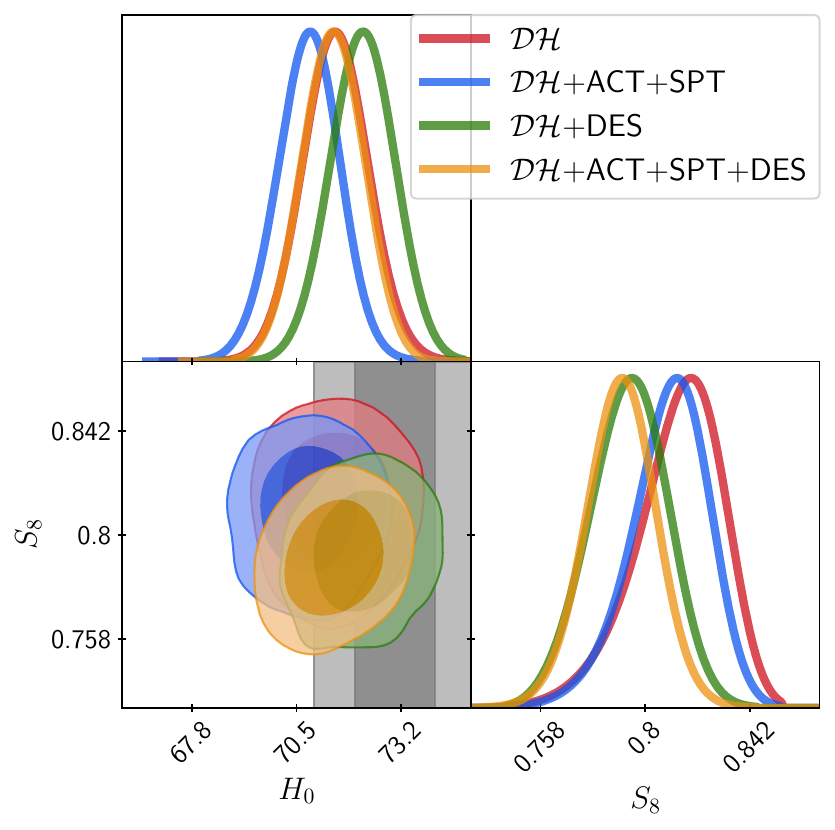}
    \caption{Posterior distribution in the $H_0$-$S_8$ plane for WZDR+, with the SH0ES prior added to each of the dataset fitted in Fig.~\ref{fig:D_s8}.}
    \label{fig:addingPriors}
\end{figure}

At the same time, it is worth noting that the inclusion of a SH0ES prior does not impact WZDR+'s ability to address the $S_8$ tension. That is, including a SH0ES prior, which we show in Fig. \ref{fig:addingPriors}, does, indeed, shift the value of $H_0$ to higher values. However, the allowed range of $S_8$ remains quite broad, and the combination of data does not significantly shift the value of $S_8$ extracted from the data without the SH0ES prior. 

In summary, we have considered the WZDR+ model in the context of a broad dataset, including recent CMB lensing results from ACT-DR6. We find the model provides a simple fit to the data, well above that of \lcdm, and naturally accommodates suppressed structure on small scales. At the present time, it does not appear to solve the Hubble tension, which is a direct result of the inclusion of primary CMB data from ACT and SPT.

Beyond simply addressing tensions, WZDR+ allows us a concrete model, grounded in well-understood particle physics to study modifications beyond \lcdm. With a limited set of parameters, it provides a solid laboratory to search for new physics in the dark sector.

\section{Acknowledgements}
We thank Melissa Joseph, Cara Giovanetti, Jeremy Tinker and Martin Schmaltz for helpful discussions and comments. We thank Cyrille Doux and Joe DeRose for discussions of the DES results, and providing data for our DES fit. Numerical computations for this work were done using the NYU High Performance Computing (HPC) cluster. NW is supported by the NSF under award 2210498, the BSF under award 2022287, and the Simons Foundation.

\bibliographystyle{apsrev}
\bibliography{main}

\appendix
\section{Tables and Full Corner Plots}
We include here tables with mean$\pm \sigma$ and best fit values, as well as Full Corner plots from the various fits.
\begin{table*}[h!]
\centering
\begin{tabular}{|c|c|c|c|c|}
\hline
&\multicolumn{2}{c|}{$\LCDM$}& \multicolumn{2}{c|}{WZDR+}\\ \hline
                             &  $\mathcal{D}$ & $\mathcal{D}$+ACT+SPT & $\mathcal{D}$ & $\mathcal{D}$+ACT+SPT  \\ \hline
$100\theta_s$                 & 1.04189       & 1.04200                               & 1.04238       & 1.04210                             \\
$\Omega_b h^2$                & 0.02242       & 0.02237                               & 0.02260       & 0.02240                              \\
$\Omega_{\rm dm}h^2$         & 0.11952       & 0.11948                               & 0.12230       & 0.11914                              \\
$\ln{10^{10}}A_s$             & 3.054         & 3.050                               & 3.055         & 3.040                                   \\
$n_s$                         & 0.9670        & 0.9692                             & 0.9719        & 0.9711                                \\
$\tau_{\rm reio}$             & 0.0589        & 0.0546                                 & 0.0592        & 0.0502                                \\
$10^6\Gamma_0\ [1/{\rm Mpc}]$ & -             & -                                    & 0.345          & 1.21                                 \\
$\log_{10}{(z_t)}$            & -             & -                                  & 4.56          & 4.42                                \\
$N_{\rm IR}$                  & -             & -                                   & 0.154         & 0.012                            \\
\hline
$\Omega_m$                    & 0.3124        & 0.3121                        & 0.3089        & 0.3085                         \\
$\sigma_8$                    & 0.8138        & 0.8131                       & 0.8005        & 0.7701              
\\
$S_8$                         & 0.8304        & 0.8293                         & 0.8123        & 0.7810                        \\ 
$H_0\ [{\rm km/s/Mpc}]$       & 67.56         & 67.57                          & 68.64         & 67.88                     \\
\hline
$\chi^2_{\rm tot}$            & 3811.52       & 5935.5                     & 3810.78       & 5931.58                          \\
$\Delta$AIC                   & -             & -                              & +5.26          & +2.08                          \\ \hline
\end{tabular}
\caption{Best-fit values for fits to the DES excluded datasets $\DD$ and $\DD$+ACT+SPT.}
\label{tab:bf}
\end{table*}

\begin{table*}[h!]
\centering
\begin{tabular}{|c|c|c|c|c|}
\hline
&\multicolumn{2}{c|}{$\LCDM$}& \multicolumn{2}{c|}{WZDR+}\\ \hline
                             &  $\mathcal{D}$+DES & $\mathcal{D}$+ACT+SPT+DES & $\mathcal{D}$+DES & $\mathcal{D}$+ACT+SPT+DES  \\ \hline
$100\theta_s$                 & 1.04193  & 1.04203                   & 1.04336 & 1.04233                   \\
$\Omega_b h^2$                &  0.02258 & 0.02243                   & 0.02285 & 0.02246                   \\
$\Omega_{\rm dm}h^2$         & 0.11835 & 0.11849                   & 0.12958 & 0.11961                   \\
$\ln{10^{10}}A_s$             &  3.043  & 3.038                     & 3.055 & 3.042                     \\
$n_s$                         & 0.9688 & 0.9690                  & 0.9844 & 0.9728                    \\
$\tau_{\rm reio}$             &  0.0564  & 0.0507                    & 0.0573 & 0.0507                    \\
$10^6\Gamma_0\ [1/{\rm Mpc}]$ & - & -                       & 0.586 & 1.15                      \\
$\log_{10}{(z_t)}$            & - & -                         & 4.35 & 4.47                      \\
$N_{\rm IR}$                  & - & -                         & 0.569 & 0.030                     \\
\hline
$\Omega_m$                    & 0.3049 & 0.3063                    &  0.3027 & 0.3083                    \\
$\sigma_8$                    & 0.8054  & 0.8046                    &  0.7882 & 0.7623      
\\
$S_8$                         & 0.8120 & 0.8130                    & 0.7917 & 0.7728                    \\ 
$H_0\ [{\rm km/s/Mpc}]$       & 68.14 & 67.99                     &  71.10 & 68.04                     \\
\hline
$\chi^2_{\rm tot}$            &  3851.14  & 5976.5                    &  3826.3  & 5953.92                   \\
$\Delta$AIC                   & - & -                         &    -18.84    & -16.58                    \\ \hline
\end{tabular}
\caption{Best-fit values for fits to the DES included datasets $\DD$+DES and $\DD$+ACT+SPT+DES.}
\label{tab:bf2}
\end{table*}

\begin{table*}[h!]
\centering
\begin{tabular}{|c|c|c|c|c|}
\hline
&\multicolumn{2}{c|}{$\LCDM$}& \multicolumn{2}{c|}{WZDR+}\\ \hline
                             &  $\mathcal{D}$ & $\mathcal{D}$+ACT+SPT & $\mathcal{D}$ & $\mathcal{D}$+ACT+SPT  \\ \hline
$100\theta_s$                 & $1.042_{-0.00029}^{+0.00029}$  &
$1.042_{-0.00025}^{+0.00025}$  & 
$1.043_{-0.00045}^{+0.00045}$ & 
$1.042_{-0.00033}^{+0.00034}$  \\

$\Omega_b h^2$                & 
$0.02239_{-0.00014}^{+0.00013}$ &
$0.02233_{-0.00011}^{+0.00011}$  & 
$0.02258_{-0.00017}^{+0.00018}$ & 
$0.02243_{-0.00013}^{+0.00013}$  \\

$\Omega_{\rm dm}h^2$         & 
$0.1195_{-0.00089}^{+0.0009}$ &
$0.1195_{-0.00085}^{+0.00084}$  & 
$0.1241_{-0.0035}^{+0.0035}$ & 
$0.1217_{-0.0022}^{+0.0023}$  \\

$\ln{10^{10}}A_s$             & 
$3.054_{-0.014}^{+0.014}$ & 
$3.053_{-0.013}^{+0.013}$  & 
$3.052_{-0.015}^{+0.015}$ & 
$3.048_{-0.015}^{+0.016}$  \\

$n_s$                         & 
$0.9657_{-0.0036}^{+0.0036}$ & 
$0.9678_{-0.0032}^{+0.0033}$ & 
$0.9739_{-0.0058}^{+0.0058}$ & 
$0.9725_{-0.0045}^{+0.0045}$   \\

$\tau_{\rm reio}$             & 
$0.05899_{-0.0074}^{+0.0074}$ & 
$0.05574_{-0.0071}^{+0.0071}$  & 
$0.05719_{-0.0077}^{+0.0078}$ & 
$0.05237_{-0.0075}^{+0.0078}$  \\

$10^6\Gamma_0\ [1/{\rm Mpc}]$ & 
- & 
- & 
$<1.766$ & 
$<2.269$  \\

$\log_{10}{(z_t)}$            & 
- & 
- & 
$4.383_{-0.16}^{+0.14}$ & 
$4.352_{-0.16}^{+0.15}$ \\

$N_{\rm IR}$                  & 
- & 
- & 
$<0.6748$ & 
$<0.4293$ \\

\hline
$\Omega_m$                    & 
$0.3123_{-0.0054}^{+0.0054}$ & 
$0.3125_{-0.005}^{+0.0051}$  & 
$0.3088_{-0.006}^{+0.0061}$ & 
$0.3114_{-0.0056}^{+0.0056}$  \\

$\sigma_8$                    & 
$0.8132_{-0.0057}^{+0.0058}$ & 
$0.8139_{-0.0054}^{+0.0054}$  & 
$0.7941_{-0.025}^{+0.023}$ & 
$0.7818_{-0.025}^{+0.025}$  \\ 

$S_8$                         & 
$0.8297_{-0.0096}^{+0.0097}$ & 
$0.8306_{-0.0092}^{+0.0093}$ & 
$0.8056_{-0.025}^{+0.023}$ & 
$0.7965_{-0.027}^{+0.025}$  \\ 

$H_0\ [{\rm km/s/Mpc}]$       & 
$67.56_{-0.4}^{+0.4}$ & 
$67.54_{-0.36}^{+0.36}$  & 
$69.07_{-1.1}^{+1.2}$ & 
$68.19_{-0.67}^{+0.68}$  \\

\hline
\end{tabular}
\caption{Mean $\pm 1\sigma$ values for fits to the DES excluded datasets $\DD$ and $\DD$+ACT+SPT. For one sided posterior we report the 95\% upper bound.}
\label{tab:meansigma}
\end{table*}
\begin{table*}[h!]
\centering
\begin{tabular}{|c|c|c|c|c|}
\hline
&\multicolumn{2}{c|}{$\LCDM$}& \multicolumn{2}{c|}{WZDR+}\\ \hline
                             &  $\mathcal{D}$+DES & $\mathcal{D}$+ACT+SPT+DES & $\mathcal{D}$+DES & $\mathcal{D}$+ACT+SPT+DES  \\ \hline
$100\theta_s$                 & 
$1.042_{-0.00029}^{+0.00029}$ & 
$1.042_{-0.00025}^{+0.00025}$ & 
$1.043_{-0.00049}^{+0.00048}$ & 
$1.043_{-0.00049}^{+0.00049}$  \\

$\Omega_b h^2$                & 
$0.02253_{-0.00013}^{+0.00013}$ & 
$0.02243_{-0.00011}^{+0.00011}$ & 
$0.02281_{-0.00018}^{+0.00017}$ & 
$0.02257_{-0.00016}^{+0.00016}$  \\

$\Omega_{\rm dm}h^2$         & 
$0.1180_{-0.00084}^{+0.00084}$ & 
$0.1182_{-0.00081}^{+0.0008}$ & 
$0.1289_{-0.0043}^{+0.0042}$ & 
$0.1239_{-0.0043}^{+0.0041}$  \\

$\ln{10^{10}}A_s$             & 
$3.045_{-0.014}^{+0.014}$ & 
$3.044_{-0.013}^{+0.013}$ & 
$3.052_{-0.015}^{+0.015}$ & 
$3.048_{-0.016}^{+0.015}$  \\

$n_s$                         & 
$0.9680_{-0.0037}^{+0.0037}$ & 
$0.9697_{-0.0033}^{+0.0033}$ & 
$0.9830_{-0.0061}^{+0.0061}$ & 
$0.9786_{-0.0066}^{+0.0065}$  \\

$\tau_{\rm reio}$             & 
$0.05648_{-0.0072}^{+0.0072}$ & 
$0.05302_{-0.007}^{+0.007}$ & 
$0.05643_{-0.0073}^{+0.0071}$ & 
$0.05174_{-0.0074}^{+0.0073}$  \\

$10^6\Gamma_0\ [1/{\rm Mpc}]$ & 
- & 
- & 
$0.6887_{-0.3}^{+0.3}$ & 
$0.8597_{-0.37}^{+0.41}$  \\

$\log_{10}{(z_t)}$            & 
- & 
- & 
$4.374_{-0.11}^{+0.11}$ & 
$4.341_{-0.13}^{+0.14}$ \\

$N_{\rm IR}$                  & 
- & 
- & 
$0.535_{-0.23}^{+0.25}$ & 
$<0.6705$  \\

\hline
$\Omega_m$                    & 
$0.3032_{-0.005}^{+0.005}$ & 
$0.3043_{-0.0047}^{+0.0047}$ & 
$0.303_{-0.0061}^{+0.006}$ & 
$0.3073_{-0.0055}^{+0.0054}$  \\

$\sigma_8$                    & 
$0.8049_{-0.0055}^{+0.0055}$ & 
$0.8060_{-0.0052}^{+0.0052}$ & 
$0.7834_{-0.02}^{+0.019}$ & 
$0.7686_{-0.018}^{+0.018}$  \\ 

$S_8$                         & 
$0.8092_{-0.0088}^{+0.0089}$ & 
$0.8117_{-0.0087}^{+0.0085}$ & 
$0.7872_{-0.018}^{+0.017}$ & 
$0.7778_{-0.018}^{+0.017}$  \\ 

$H_0\ [{\rm km/s/Mpc}]$       & 
$68.25_{-0.38}^{+0.38}$ & 
$68.13_{-0.35}^{+0.35}$ & 
$70.9_{-1.4}^{+1.4}$ & 
$69.2_{-1.2}^{+1.2}$  \\

\hline
\end{tabular}
\caption{Mean $\pm 1\sigma$ values for fits to the DES included datasets $\DD$+DES and $\DD$+ACT+SPT+DES. For one sided posterior we report the 95\% upper bound.}
\label{tab:meansigma2}
\end{table*}

\begin{figure*}
    \centering
    \includegraphics[width=\linewidth]{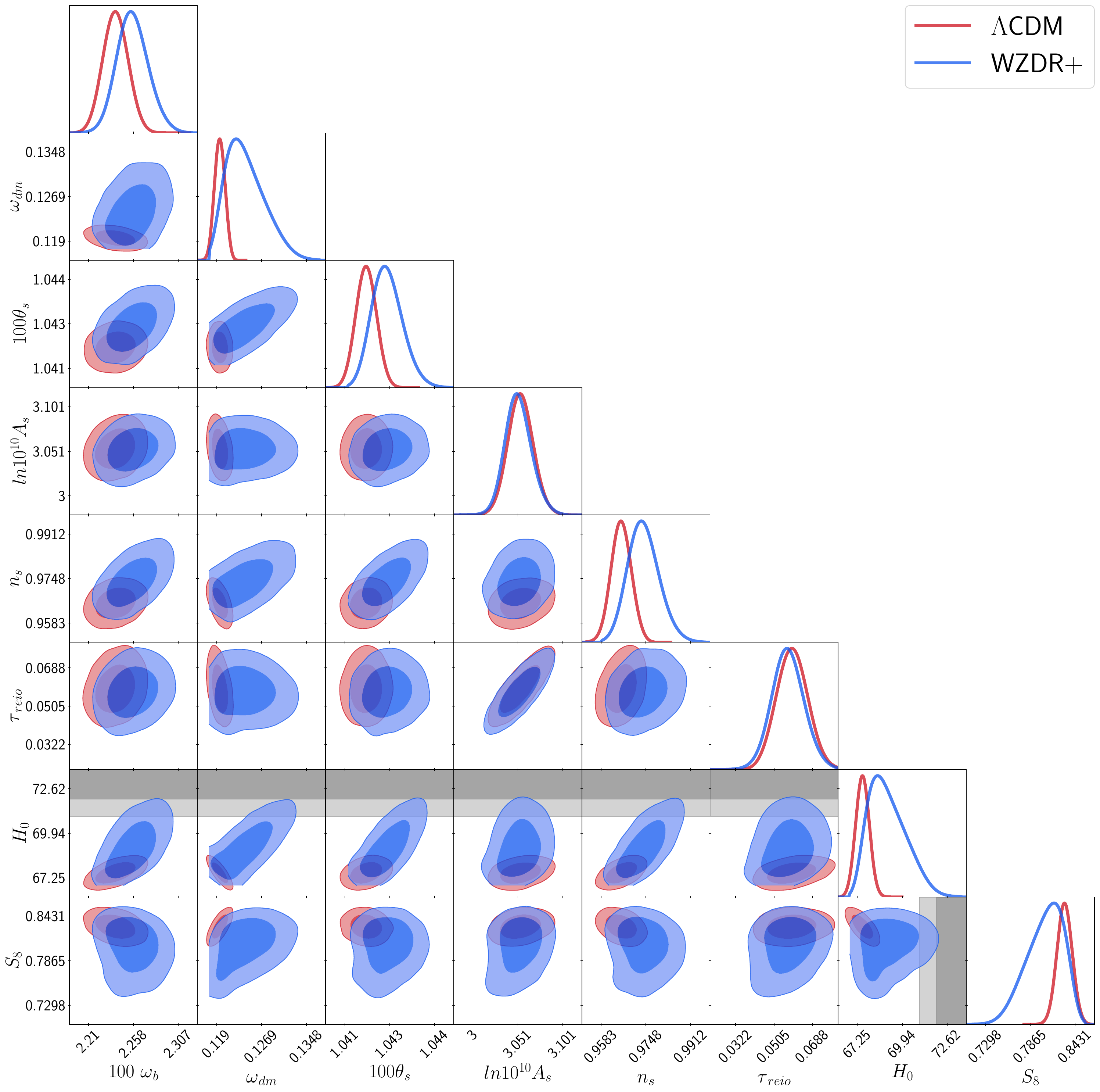}
    \caption{Comparison of the posterior distributions for LCDM and WZDR+ fitted to $\DD$. The grey band shows the $1\sigma$ and $2\sigma$ regions of the SH0ES measurement.}
    \label{fig:enter-label}
\end{figure*}

\begin{figure*}
    \centering
    \includegraphics[width=\linewidth]{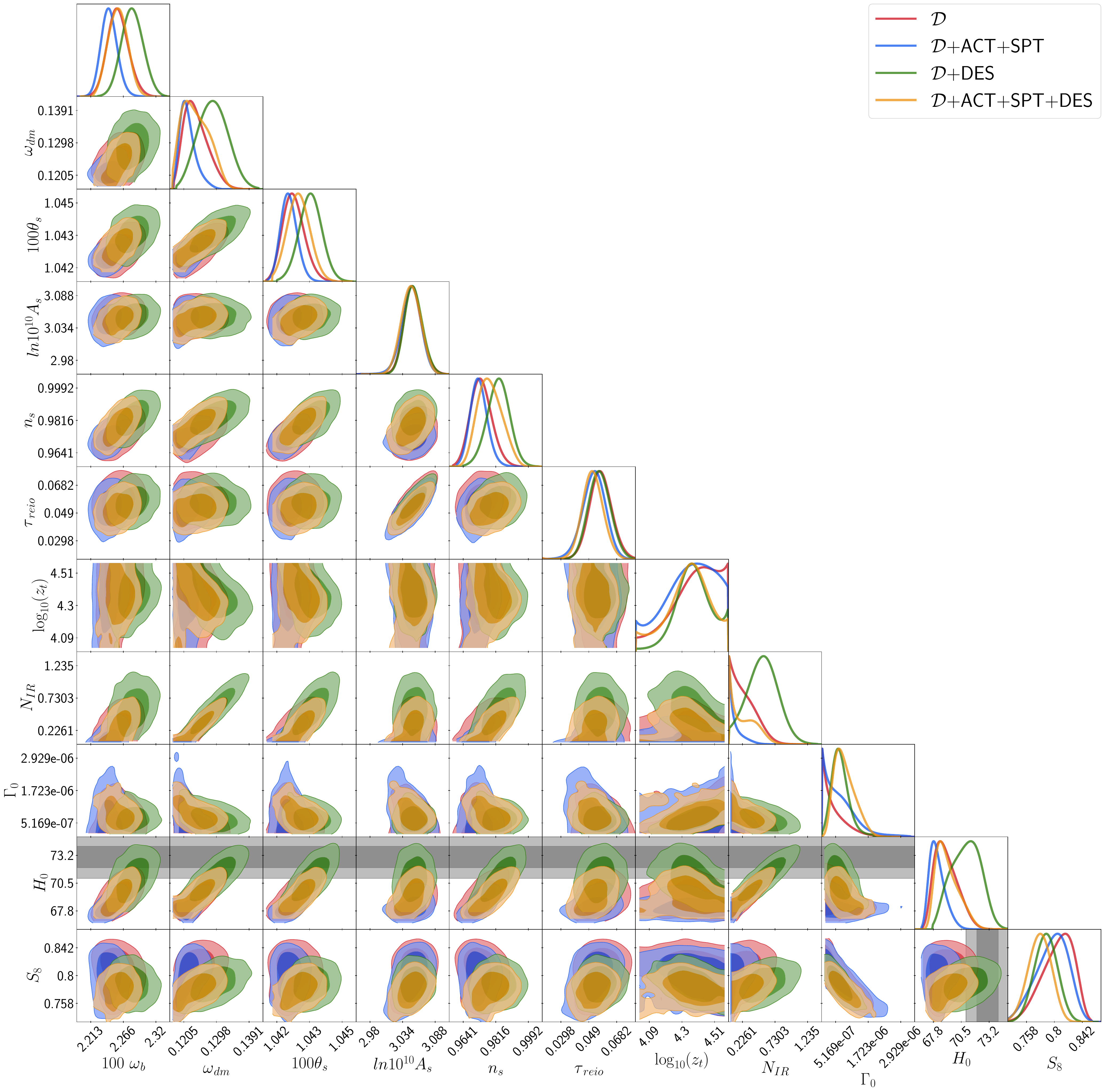}
    \caption{Posterior distrubtions for WZDR+ fitted to various datasets without SH0ES. Like in \cite{joseph}, we find a preference for the transition redshift $\log_{10}{(z_t)}\sim4.3$ even with only CMB data.}
    \label{fig:enter-label}
\end{figure*}

\begin{figure*}
    \centering
    \includegraphics[width=\linewidth]{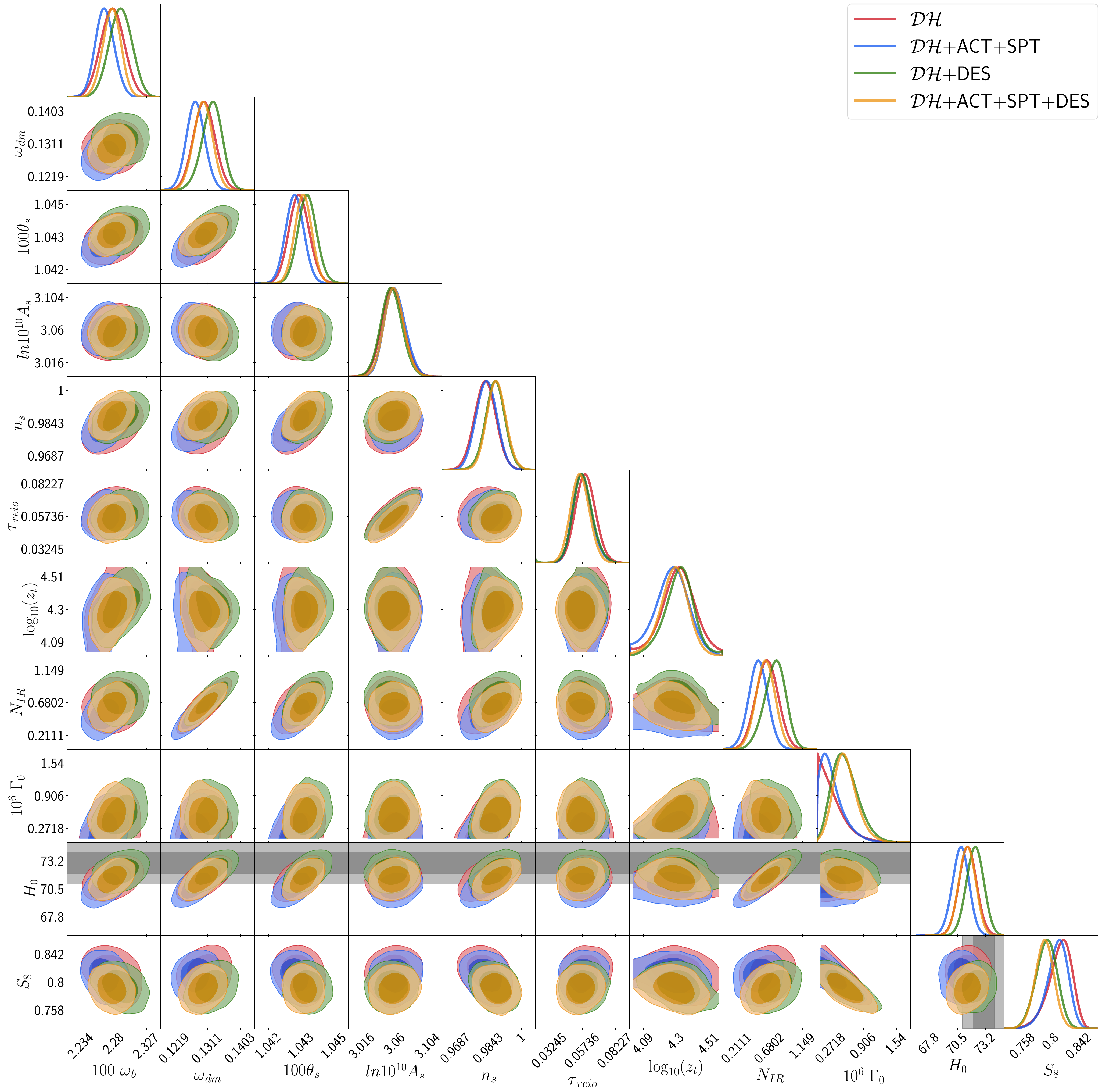}
    \caption{Posterior distribution of WZDR+ fitted to a various datasets with SH0ES.}
    \label{fig:enter-label}
\end{figure*}

\end{document}